\documentclass[a4paper,10pt]{article}

\usepackage{xcolor}

\usepackage{jheppub} 

\usepackage{amsmath,amssymb}

\usepackage[compat=1.1.0]{tikz-feynman}
\usepackage{tikz}
\usepackage{mathtools}

\usepackage{simplewick}
\usepackage{graphicx}

\title{Interacting Fractons in 2+1-Dimensional Quantum Field Theory}

\preprint{UTTG-25-2021}

\author[a]{Jacques Distler,}
\author[b]{Murtaza Jafry,}
\author[a]{Andreas Karch,}
\author[a]{Amir Raz}

\affiliation[a]{University of Texas, Austin, Physics Department, Austin TX 78712, USA}
\affiliation[b]{Department of Physics, University of Washington, Seattle, WA, 98195-1560, USA}

\emailAdd{distler@golem.ph.utexas.edu}
\emailAdd{jafrym@uw.edu}
\emailAdd{karcha@utexas.edu}
\emailAdd{araz@utexas.edu}

\abstract{We analyze, in perturbation theory, a theory of weakly interacting fractons and non-relativistic fermions in a 2+1 dimensional Quantum Field Theory. In particular we compute the 1-loop corrections to the self energies and interaction vertex, and calculate the associated 1-loop Renormalization Group flows of the coupling constants. Surprisingly, we find that the fracton-fermion coupling does not flow due to an emergent coordinate-dependent symmetry of the effective Lagrangian, making this model a well-defined quantum field theory. We provide additional discussions on the regularization and renormalization of interacting fractonic theories, as well as both qualitative and quantitative remarks regarding the theory at finite temperature and finite chemical potential. }

\begin{document} 

\newcommand{\eq}{equation}
\newcommand{\pa}{\partial}
\newcommand{\da}{\dagger}
\newcommand{\M}{\mathcal{M}}
\newcommand{\Tr}{\text{Tr}}
\newcommand{\s}{\slashed}
\newcommand{\fr}{\frac}
\newcommand{\La}{\mathcal{L}}
\newcommand{\ep}{\epsilon}
\newcommand{\dl}{\displaylimits}

\maketitle
\flushbottom

\section{Introduction} 
Field theories of fractons are characterized by a rich set of symmetries \cite{Nan,Pret}. In many of the simplest examples, these symmetries are indeed strong enough to ensure that the resulting field theory is free at low energies. All interaction terms consistent with symmetries are irrelevant in the sense of the renormalization group. Many interesting field theories of this type involving a single real scalar which, in an abuse of nomenclature we will simply refer to as a fracton, have been analyzed in detail in \cite{paramekanti2002ring,Seiberg_2020,sei,Seiberg:2020wsg}. Interacting theories with these symmetries can be constructed by adding extra fields; one very simple option being to promote the scalar to a complex scalar. While no longer free, these theories are often intractable. One complication that arises for example in the case of the complex scalars are kinetic terms that are quartic in the field \cite{Pretko_2018,Gromov2019}, making the theory difficult to analyze with traditional perturbative techniques (see for example \cite{Peng2021}.) Calculable models of interacting fractons are hard to come by.

In this paper, we will present one such interacting fracton model with a controlled perturbative expansion and will calculate its properties to leading non-trivial order. The model describes the interaction of a non-relativistic fermion with a fracton scalar. This model will have a momentum dipole symmetry given by the shift of the fracton scalar $\phi \rightarrow \phi + f(x) + g(y)$, for arbitrary functions $f$ and $g$. One can deform this theory by introducing a non-relativistic fermion $\psi$ that couples with the scalar via the interaction $\lambda \psi^\da \psi\pa_x \pa_y \phi$. This interaction is marginal and both preserves the dipole symmetry as well as the $\mathbb{Z}_4$ rotational symmetry. In fact these symmetries exclude any additional relevant or marginal interactions. This allows us to study the theory using conventional perturbation theory. In particular, we calculate the $\beta$-functions of the theory to 1-loop order. Interestingly we find that the theory has a vanishing beta function, and so describes a well-defined quantum field theory in its own right. We also briefly analyze the vacuum structure of the theory.

The paper is organized as follows. We first introduce the lattice construction of the interacting fermion and scalar in section \ref{sec:2}. We then present the formal Lagrangian for the continuum field theory in section \ref{sec:3}, followed by a description of the perturbative framework. This includes describing the associated Feynman rules, counter-terms, and regularization and renormalization schemes. In section \ref{sec:4} we calculate all the divergent diagrams of the theory at 1-loop order, and provide the counter-terms to subtract of the associated UV divergences. In section \ref{sec:5} we calculate the associated beta functions for the associated couplings constants for the theory, and show that the $\beta$ function for the coupling $\lambda$ vanishes due to an emergent symmetry of the theory. In section \ref{sec:7} we discuss the theory at non-zero temperature, while in section \ref{sec:8} we analyze the theory when the fermions have a finite density in  the ground state. Finally, we conclude with a summary in section \ref{sec:9}. 
\section{Lattice model} \label{sec:2} 

The field theory we wish to discuss can be considered to be the low energy limit of a lattice Hamiltonian of the XY-plaquette model \cite{paramekanti2002ring,sei} interacting with a fermionic field. This lattice theory consists of a compact scalar $\phi$ living on each site $s$ of the lattice, with an XY-plaquette model Hamiltonian \cite{paramekanti2002ring,sei}, 
\begin{\eq}
H_{scal} = \fr{u}{2} \sum_s \pi_s^2 - K \sum_s \cos(\Delta_{xy}  \phi_s) ,
\end{\eq}
where $\Delta_{xy} \phi_{\hat{x},\hat{y}} = \phi_{\hat{x}+1,\hat{y}+1} -\phi_{\hat{x}+1,\hat{y}} - \phi_{\hat{x},\hat{y}+1} + \phi_{\hat{x},\hat{y}}$. Here $\pi_s$ are the conjugate momentum modes for the scalar field. This Hamiltonian is invariant under shifting all the scalars fields $\phi$ on a specific coordinate line of the lattice, and also under the $\mathbb{Z}_4$ rotations of the lattice.

In addition we introduce a fermionic field, $\psi$, to each lattice site. Though we can consider these fermions to have some spin (usually spin $1/2$,) the spin operators will always act as an internal symmetry of the model (as is the case in non-relativistic theories,) so for simplicity we focus on a spinless fermion. The Hamiltonian for these fermions is taken to be a standard nearest neighbor interaction,
\begin{\eq}
H_{fer} = - J \sum_{\left<s,s'\right>}\psi^\da_s \psi_{s'}
+ h \sum_{s} \psi^\da_s \psi_s,
\end{\eq}
which in the continuum can becomes a free non-relativistic fermion. The operators obey the canonical commutation and anti-commutation relations of the form $[\phi_s,\pi_{s'}] = i\delta_{s,s'}$ and $\{\psi_n,\psi^\da_{n'}\} = i\delta_{n,n'}$. 

We would like to couple the XY-plaquette scalar to the fermions in a way that preserves the subsystem symmetries of the scalar field. The simplest local interaction term that meets this criterion is
\begin{\eq}
H_{int} = R \sum_s \psi^\da_s \psi_s \sin(\Delta_{xy} \phi_s) .
\end{\eq}
This term also requires that the $\phi$ field transform in the spin-2 representation the $\mathbb{Z}_4$ rotation symmetry in order to stay invariant. From this, we can write the complete lattice Hamiltonian as
\begin{\eq}
\begin{split}
H &= H_{scal} + H_{fer}  + H_{int} \\ &= \fr{u}{2} \sum_s \pi_s^2 - K \sum_s \cos(\Delta_{xy} \phi_s)  - J \sum_{\left<s,s'\right>}\psi^\da_s \psi_{s'} 
+ h \sum_{s} \psi^\da_s \psi_s 
+R\sum_n \psi^\da_n \psi_n \sin(\Delta_{xy} \phi) 
\end{split}
\end{\eq}
Though this lattice model is interesting in its own right (or perhaps similar but more solvable lattice models inspired by \cite{gorantla2021modified,gorantla2021lowenergy},) we will focus on its continuum field theory description, formulated in the next section.

\section{Continuum field theory} \label{sec:3} 

We are interested in studying the continuum limit of the lattice model considered in the previous section. The continuum limit of the XY-plaquette model was carefully constructed in \cite{sei}, and consists of a single compact scalar field $\phi$ with the Lagrangian
\begin{\eq} \label{eq:lag_scal}
\La_{scal} = \fr{\mu_0}{2}(\pa_0\phi)^2 - \fr{1}{2\mu}(\pa_x\pa_y \phi)^2 .
\end{\eq}
The shift symmetry on the lattice becomes the subsystem symmetry $\phi(x,y) \rightarrow \phi(x,y) + f_x(x) + f_y(y)$ for arbitrary function $f_x$ and $f_y$.

The fermionic lattice Hamiltonian also has a well known continuum limit of a free non-relativistic fermion $\psi$ with the action
\begin{\eq} \label{eq:lag_ferm}
\La_{fer} = \psi^\da \left( i\pa_0 +  \fr{\nabla^2}{2m}  - \gamma \right)  \psi ,
\end{\eq}
where $m$ is the mass of the fermion and $-\gamma$ is the chemical potential of the fermion.

It remains to understand how the lattice interaction between the fermion and the scalar will look within the continuum theory. Naively we can expand $\sin(\Delta_{xy} \phi) \approx \Delta_{xy} \phi + O(\Delta_{xy}^2\phi)$, so we expect the leading order interaction to be
\begin{\eq} \label{eq:interaction}
\La_{int} = \lambda \psi^\da \psi \pa_x\pa_y \phi .
\end{\eq}
As in the lattice model, invariance of the interaction term requires that the scalar, $\phi$, transforms in the spin $2$ representation of the discrete $\mathbb{Z}_4$ rotations. For comparison, this scalar field $\phi$ is the equivalent to the $\phi^{xy}$ scalar field in \cite{sei}.

To verify that this is in fact the interaction term in the continuum theory we must verify that this is the only relevant or marginal interaction term that respects the symmetries in the continuum description. To work out the scaling dimensions, first note that both the scalar and the fermion kinetic terms have twice as many spatial derivatives as they have temporal derivatives, so the free theory has a scale symmetry with dynamical critical exponent $z=2$ (as is standard in non-relativistic theories) and so the derivatives have scaling dimensions
\begin{\eq}
[\partial_0]=2, \quad [\partial_x]=[\partial_y]=1.
\end{\eq}
The kinetic terms then fix the scaling dimensions of the fields to be
\begin{\eq}
[\psi] = 1, \;\; \;  [\phi] = 0 
\end{\eq}
leading to a marginal coupling for the interaction term, that is $[\lambda]=0$. Any additional interaction term that respects the subsystem symmetry would have additional powers of $\partial_x \partial_y \phi$ or $\psi^\da \psi$, and so would be irrelevant. Thus \eqref{eq:interaction} is the only possible marginal or relevant interaction term in the continuum theory.

From this, we find that the full continuum theory, including all relevant and marginal terms consistent with symmetry, in \textbf{Euclidean Signature} is
\begin{\eq} \label{eq:Lag_euc}
\begin{split}
\La &= \La_{scal} + \La_{fer} + \La_{int} \\ &= \fr{\mu_0}{2}(\pa_0\phi)^2 + \fr{1}{2\mu}(\pa_x\pa_y \phi)^2  + \psi^\da \bigg ( \pa_0 -  \fr{\nabla^2}{2m} \bigg  ) \psi - \lambda \psi^\da \psi \pa_x\pa_y \phi  + \gamma \psi^\da \psi .
\end{split} 
\end{\eq}
One can also write down the theory in \textbf{Minkowski Signature},
\begin{\eq}\label{MinkLagrangian}
\begin{split}
\La &= \La_{scal} + \La_{fer} + \La_{int} \\ &= \fr{\mu_0}{2}(\pa_0\phi)^2 - \fr{1}{2\mu}(\pa_x\pa_y \phi)^2  + \psi^\da \bigg ( i\pa_0 +  \fr{\nabla^2}{2m} \bigg  ) \psi +  \lambda \psi^\da \psi \pa_x\pa_y \phi - \gamma \psi^\da \psi .
\end{split} 
\end{\eq}

We will study this interacting theory using conventional perturbation theory techniques, assuming the marginal interaction coupling $\lambda$ is small. In the next subsections we further develop the perturbative framework of this model, including stating the Feynman rules and the regularization scheme we use to define the (divergent part of the) counter-terms.
 
\subsection{Feynman rules} 

The Feynman rules for this theory are derived from the free Lagrangians, namely \eqref{eq:lag_scal} and \eqref{eq:lag_ferm}. In position space the (free) fermionic propagator is
\begin{\eq}
\contraction{}{\psi^\da}{(\tau,x,y) }{\psi} \psi^\da(\tau,x,y) \psi(0)  = \fr{1}{(2\pi)^3}\int_{-\infty}^{\infty}  d\omega dk_xdk_y \fr{e^{-i\omega \tau - ik_x x- ik_y y}}{-i\omega - \fr{k_x^2+k_y^2}{2m}} ,
\end{\eq}
while the (free) propagator for the scalar field is 
\begin{\eq}
\contraction{}{\phi}{(\tau,x,y) }{\phi} \phi(\tau,x,y) \phi(0)= \fr{1}{(2\pi)^3} \int_{-\infty}^{\infty} d\omega dk_xdk_y \fr{e^{i\omega \tau + ik_x x+ ik_y y}}{\mu_0 \omega^2 + \fr{k_x^2k_y^2}{\mu}} .
\end{\eq}

Adding the interaction vertex, we can write the Feynman rules in momentum space and in \textbf{Euclidean Signature} as
\begin{\eq}
\begin{split}
\text{Vertex} &= -\lambda k_xk_y , \\ 
\contraction{}{\psi^\da}{(\omega,k_x,k_y)}{ \psi} \psi^\da(\omega,k_x,k_y) \psi(-\omega,-k_x,-k_y)  &= \fr{1}{-i\omega + \fr{k_x^2+k_y^2}{2m} + \gamma } ,\\ 
\contraction{}{\phi}{(\omega,k_x,k_y)}{ \phi} \phi(\omega,k_x,k_y) \phi(-\omega,-k_x,-k_y)  &= \fr{1}{\mu_0 \omega^2 + \fr{k_x^2k_y^2}{\mu}}.
\end{split}
\end{\eq}
The corresponding  Feynman rules in \textbf{Minkowski Signature} are
\begin{\eq}
\begin{split}
\text{Vertex} &= i\lambda k_xk_y \\ 
\contraction{}{\psi^\da}{(\omega,k_x,k_y)}{ \psi} \psi^\da(\omega,k_x,k_y) \psi(-\omega,-k_x,-k_y)  &= \fr{1}{\omega - \fr{k_x^2+k_y^2}{2m} + \gamma - i\ep} \\ \contraction{}{\phi}{(\omega,k_x,k_y)}{ \phi} \phi(\omega,k_x,k_y) \phi(-\omega,-k_x,-k_y)  &= \fr{1}{-\mu_0 \omega^2 + \fr{k_x^2k_y^2}{\mu} + i\ep}
\end{split}
\end{\eq}
One can see that under a Wick rotation of the Minkowski signature, $t \rightarrow -i\tau$, one will reproduce the Euclidean propagators from the Minkowski propagators.

\subsection{Counter-term definitions} \label{sec:counterterms}
To renormalize this theory, we introduce local counterterms and subtract the UV divergences using a rotationally-invariant hard momentum cutoff. The full Euclidean Lagrangian with these counterterms reads 
\begin{\eq} \label{eq:full_lag}
\begin{aligned}
    \La = & \fr{\mu_0}{2}(\pa_0\phi)^2 + \fr{1}{2\mu}(\pa_x\pa_y \phi)^2  + \psi^\da \bigg ( \pa_0 -  \fr{\nabla^2}{2m} \bigg  ) \psi - \lambda \psi^\da \psi \pa_x\pa_y \phi  + \gamma \psi^\da \psi \\ 
    &+ \frac{\delta_{\mu_0}}{2}(\pa_0 \phi)^2 
    + \fr{\delta_{1/\mu}}{2}  (\pa_x \pa_y \phi)^2
    + \delta_{Z_\psi} \psi^\da ( \pa_0  ) \psi 
    - \frac{\delta_{1/m}}{2} \psi^\da \nabla^2 \psi 
    - \delta_\lambda \psi^\da \psi \pa_x \pa_y \phi  + \delta_\gamma \psi^\da \psi 
\end{aligned}
\end{\eq} 
The form and signs for the counterterms follows the convention of \cite{Peskin}. 

Though \eqref{eq:full_lag} is the most general renormalized Lagrangian, we can show that in this theory $\delta_{\mu_0} = 0$ to all orders in perturbation theory using standard power counting of divergences. Indeed any diagram contributing to the 1PI scalar 2-point function must include a fermion loop that the external scalars connect to, similar to the 1-loop diagram in figure \ref{fig:1}. Then the nature of the vertex interaction implies such a diagram is proportional to $k_x^2 k_y^2$, where $k_{x,y}$ are the external spatial  momenta. Then, by power counting, we see that such diagrams are at most logarithmically divergent, implying that they only contribute to $\delta_{1/\mu}$, and so $\delta_{\mu_0} = 0$ to all orders in $\lambda$.  It turns out that at 1-loop order $\delta_{1/\mu} = 0$ as well, due to the vanishing of the fermion loop integral, which will be shown in section \ref{sec:scalarloop}.

We note that one can introduce an alternative marginal interaction term
\begin{equation}
    \lambda'\psi^\dagger \psi \partial_0 \phi
\end{equation}
which would lead to a nonzero value for $\delta_{\mu_0}$. However, this coupling is only possible if we take $\phi$ to be uncharged under the $\mathbb{Z}_4$ rotational symmetry, so we cannot include both marginal coupling terms at the same time. Since the lattice theory suggests that $\phi$ should transform in the spin-2 representation of the $\mathbb{Z}_4$ rotational symmetry, we set $\lambda'=0$.


\subsection{Regularization and renormalization scheme}

To regularize this theory we will employ a rotationally-invariant hard momentum cutoff, $\Lambda$, for the conventional renormalization group (RG) flow in momenta space. The divergences are absorbed into the local counter-terms introduced in the previous section. This procedure is the same as the standard picture of re-scaling space.

We choose to use a hard momentum cutoff as many of the other standard regularization schemes seem ill-equipped to deal with the unique dispersion relation of the scalar field. In particular it is not clear how to use dimensional regularization to regularize this field theory. We note that it may be possible to use different regulators to regulate this theory, such as a Pauli-Villars type regulator, which may have their own advantages and shortcomings. 

It is interesting to note that the standard RG flow in momenta space is not the only renormalization scheme proposed for these types of field theories. An alternative approach, suggested by \cite{lake2021rg}, involves integrating out the fast moving or high frequency modes. As the dispersion relation for the scalar is nonstandard, this modified RG scheme for a theory consisting of only the scalar field amounts to integrating only over the region in momentum space defined by $k_x^2 k_y^2  < \Lambda^2$ for some energy cutoff $\Lambda$. As our theory consists of a fermion with a standard non-relativistic kinetic term along with the scalar, integrating over the surface $k_x^2 k_y^2  < \Lambda^2$ seems unnatural as it captures many of the high energy fermion modes.

A more conventional strategy would be to implement a cutoff in the frequency $\omega$, similar to the standard RG procedures when computing the RG flow near the Fermi surface \cite{polchinski1992,Shankar1994}. Such a hard cutoff would not leave all the integrals in the theory regularized, and an additional regularization scheme would be required. However, when working at zero density (fixed non-positive chemical potential, or equivalently $\gamma \geq 0$,) we expect the two different RG prescriptions to coincide. This is due to the fact that every loop will contain at least one fermion propagator, so integrating out all the high energy states ensures that a large momenta cannot flow through the loops. 

At finite densities this standard RG prescription in momentum space is no long valid as the low energy states are near the Fermi surface \cite{polchinski1992,Shankar1994}. Due to this complication we will only work at zero density (that is we assume $\gamma \geq 0$) for the next few sections, and then make some general comments about the theory at finite density in section \ref{sec:8}.

\section{Diagrams to 1-loop order} \label{sec:4} 
There are three diagrams which are divergent to 1-loop order in this theory. The associated UV divergences are subtracted off by the counter-terms, $\{ \delta_{Z_\psi},\delta_{1/m},\delta_{1/\mu},\delta_\lambda, \delta_\gamma \}$. The diagrams will be computed in Euclidean space, $\mathbb{R}^3$.

\subsection{Scalar self-energy} \label{sec:scalarloop}
\begin{figure}
\begin{center} 
\includegraphics[width = 3.5in]{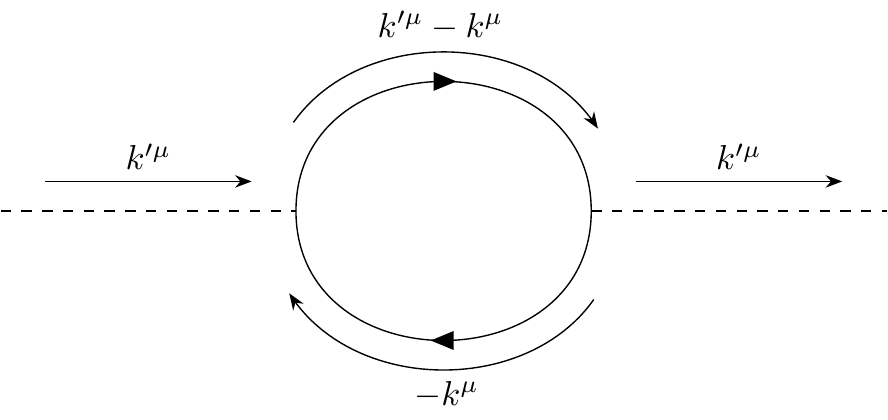}
 \end{center}
 \caption{The Feynman diagram contributing to the scalar self energy at 1-loop. In this figure, the dotted lines represent the scalar propagators while the solid lines represent the fermion propagators. The self-energy contribution here represents the only second order contribution to the 1PI diagram for the scalars within this theory.}
\label{fig:1}
 \end{figure}

The 1-loop diagram contributing to the scalar 1PI propagator is in figure \ref{fig:1}. The contribution of this diagram to the scalar self-energy is
\begin{\eq}
\begin{split}
\Sigma_\phi  &=  \fr{\lambda^2k_x'^2k_y'^2}{(2\pi)^3} \int d\omega \int d^2k \bigg ( \fr{1}{+i\omega + \fr{k^2}{2m}+\gamma} \bigg ) \bigg ( \fr{1}{-i(\omega'-\omega) + \fr{(k'-k)^2}{2m}+\gamma}  \bigg ) .
\end{split}
\end{\eq}

Notice that both $\omega$ poles of the integrand lie in the upper half plane, so we can close the $\omega$ integral contour in the lower half plane and see that $\Sigma_\phi = 0$. Thus the scalar is not renormalized, and the counter term $\delta_{1/\mu}$ to 1-loop order is simply
\begin{\eq}
\delta_{1/\mu} = 0.
\end{\eq}
In fact this is simply due to the causal structure of the retarded fermion propagator, which ensures any fermion loop will vanish. Thus the scalar self energy remains zero at all orders in perturbation theory.


\subsection{Fermion self-energy}
\begin{figure}
\begin{center} 
\includegraphics[width = 3.5in]{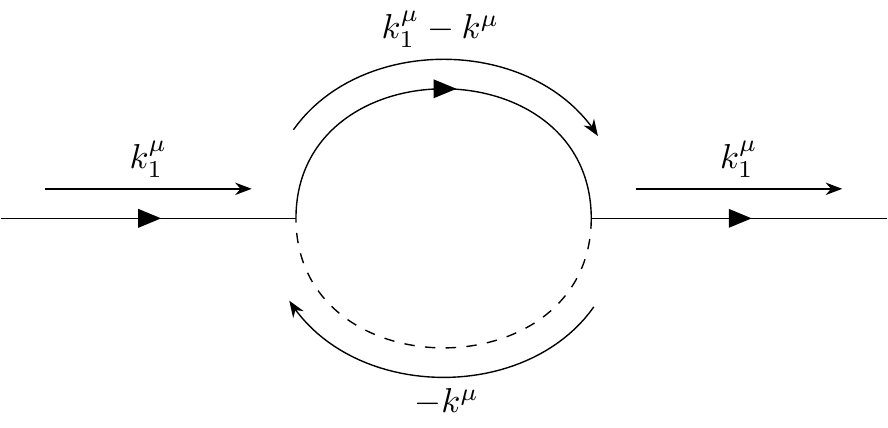}
 \end{center}
 \caption{The Feynman diagram contributing to the fermion self energy at 1-loop. In this figure, the dotted lines represent the scalar propagators while the solid lines represent the fermion propagators. The self-energy contribution here is the only second order contribution to the 1PI diagram for the fermion in this theory.}
\label{fig:2}
 \end{figure}

The 1-loop diagram contributing to the fermion 1PI propagator is in figure \ref{fig:2}. The contribution of this diagram to the fermion self energy is 
\begin{\eq} \label{eq:fermion_1PI}
\Sigma_\psi = (-\lambda)^2  \int \fr{d\omega d^2k}{(2\pi)^3} \fr{1}{-i(\omega_1 - \omega) + \fr{(k_1 - k )^2}{2m} + \gamma}\fr{k_x^2k_y^2}{\mu_0 \omega^2 + \fr{k_x^2k_y^2}{\mu}} .
\end{\eq}
This integral has three distinct UV divergences that will be absorbed by the counter-terms $\delta_{Z_\psi}$, $\delta_{1/2m}$ and $\delta_\gamma$ in equation \eqref{eq:full_lag}. This arises from the fact that the self-energy can be expressed as 
\begin{\eq} \label{eq:4.4}
\Sigma_\psi = \bigg (\fr{k_1^2}{2m} + \gamma - i\omega_1 \bigg ) \Xi_1 \log(\Lambda) + \frac{k_1^2}{2} \Xi_2 \log(\Lambda) + \fr{\Lambda^2}{2} \Xi_3,
\end{\eq}
where $\Xi_1$, $\Xi_2$, and $\Xi_3$ are three functions of the couplings which are momentum and cutoff independent. We expanded the fermion self-energy in this manner as $\Xi_1$ and $\Xi_2$ have physical significance, namely $\Xi_1$ is related to the anomalous dimension of the fermion, while $\Xi_2$ is related to the beta-function for $m$. This will be discussed in more detail in the next section.

In terms of these functions, the counter-terms $\delta_{Z_\psi}$, $\delta_{1/m}$ and $\delta_\gamma$ are
\begin{\eq} \label{eq:4.5}
\begin{split}
\delta_{Z_\psi} &=  -\Xi_1 \log(\Lambda) \\ 
\delta_{1/m} &= -\fr{\Xi_1}{m} \log (\Lambda) - \Xi_2\log(\Lambda) \\
\delta_\gamma &= -\fr{\Lambda^2}{2} \Xi_3 - \gamma \Xi_1 \log(\Lambda) 
\end{split}
\end{\eq}

Evaluating the integral in \eqref{eq:fermion_1PI} using an ultraviolet cutoff, $\Lambda$, results in the following values for $\Xi_1$, $\Xi_2$, and $\Xi_3$:
\begin{align}
\Xi_1 &=  -\lambda^2 \fr{m^2 \mu}{\pi^2} \frac{ \sqrt{\mu \mu_0}}{\mu\mu_0 - m^2}\left\{ 1 + \fr{m}{(\mu\mu_0-m^2)^{1/2}} \left( \arctan\left(\fr{m}{\sqrt{\mu \mu_0-m^2}}\right) - \fr{\pi}{2} \right)  \right\}, \\
\Xi_2 &= \lambda^2 \fr{m \mu}{\pi^2}  \fr{(\mu \mu_0)^{3/2}}{(\mu\mu_0-m^2)^2} \left\{ 2 + \fr{m^2}{\mu\mu_0} + \fr{3m}{\sqrt{\mu\mu_0-m^2}} \bigg (\arctan \bigg (\fr{m}{\sqrt{\mu\mu_0-m^2}} \bigg )  - \fr{\pi}{2} \bigg ) \right\} , \\
\Xi_3 &= \lambda^2 \fr{\mu }{2\pi^2} \left\{  \fr{\pi}{2} +  \sqrt{\fr{\mu\mu_0}{\mu\mu_0-m^2}}\left( \arctan \bigg ( \fr{m}{\sqrt{\mu\mu_0 -m^2}} \bigg ) - \fr{\pi}{2} \right)  \right\}.
\end{align}
The full explicit calculation for the Fermion 1PI diagram can be found in Appendix \ref{A2}. 

\subsection{Vertex correction} 
\begin{figure}
\begin{center}
\includegraphics[width = 2.5in]{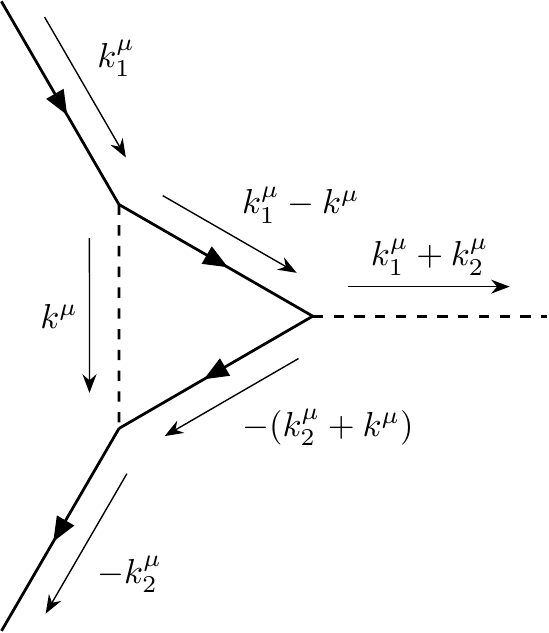}
\end{center}
\caption{Here we have the vertex correction diagram for the scalar $+$ fermion theory. Here the solid line represents the fermion, and the dashed line represents the scalar.}
\label{fig:7}
\end{figure}

The final divergent 1-loop diagram corrects the scalar--fermion vertex. The diagram for this vertex correction is in figure \ref{fig:7}. 
\begin{\eq}
\Gamma = (-\lambda)^3 \int \fr{d\omega d^2k}{(2\pi)^3} \fr{1}{-i(\omega_1 - \omega) + \fr{(k_1 - k )^2}{2m} + \gamma} \fr{k_x^2k_y^2}{\mu_0 \omega^2 + \fr{k_x^2k_y^2}{\mu}}\fr{(k_1+k_2)_x(k_1+k_2)_y}{i(\omega_2 + \omega) + \fr{(k_2 + k )^2}{2m} + \gamma} 
\end{\eq}
Again, employing a hard momentum cutoff, $\Lambda$, we find that the UV divergent term in this integral is proportional to $\log(\Lambda)$. The associated counter-term to absorb this divergence is $\delta_\lambda$ in \eqref{eq:full_lag}. This counter-term is
\begin{\eq} \label{eq:4.10} 
\delta_\lambda =\lambda^3 \fr{m^2 \mu}{\pi^2} \fr{\sqrt{\mu\mu_0}}{\mu\mu_0 - m^2}
\left\{ 1 +\fr{m}{(\mu\mu_0-m^2)^{1/2}} \left( \arctan \left( \fr{m}{\sqrt{\mu\mu_0-m^2}} \right) - \fr{\pi}{2} \right)  \right\} \log(\Lambda) 
\end{\eq}
The full explicit calculation of this Feynman diagram can be found in Appendix \ref{A3}.

\begin{figure}[h] 
\begin{center}
\includegraphics[width = \textwidth]{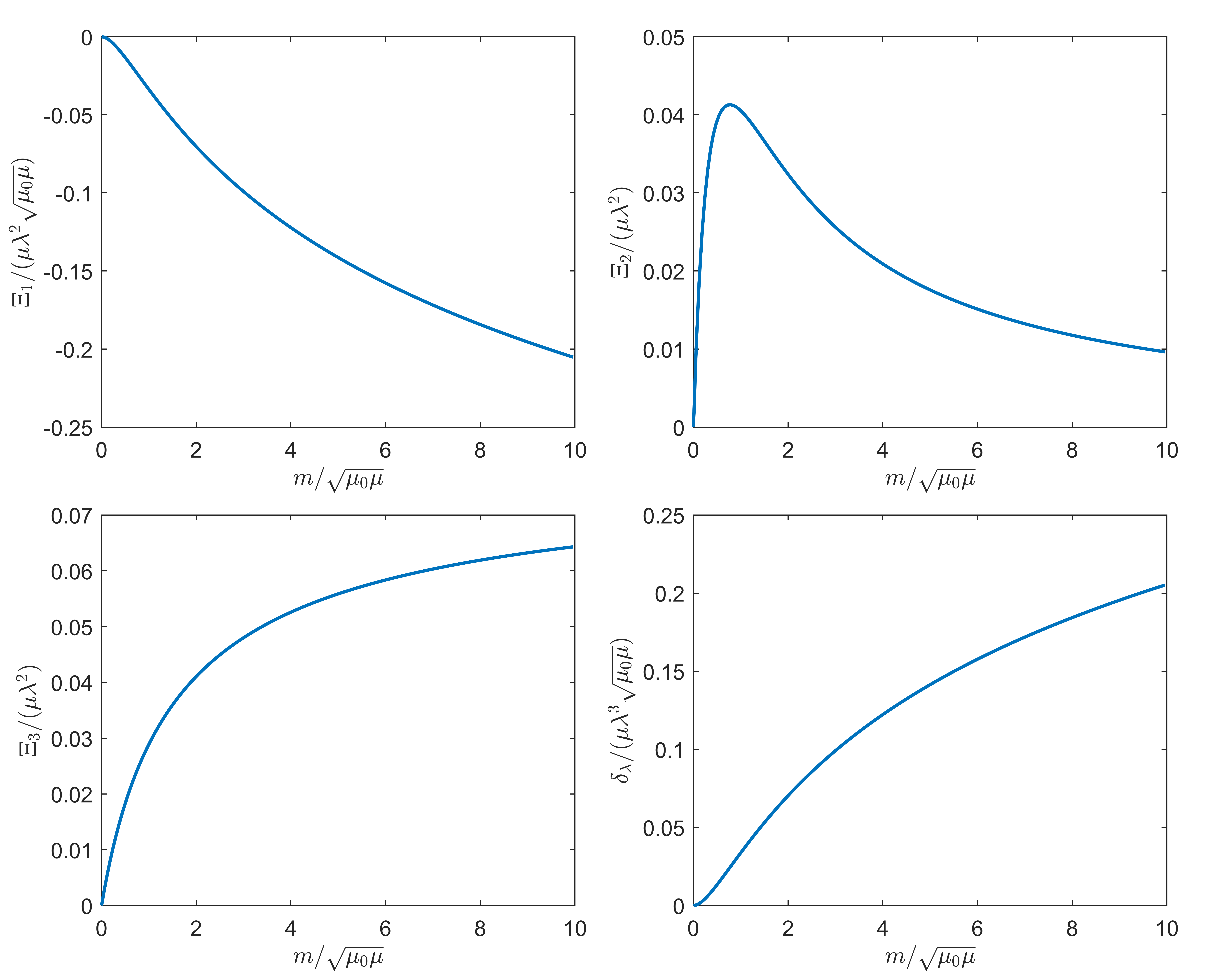}
\end{center}
\caption{Plots of the various counter-terms as a function of $m/\sqrt{\mu_0 \mu}$. Note that these counter-terms are continuous across $\mu\mu_0 = m^2$.}
\label{fig:3}
\end{figure}

Though most of the counter-terms appear to have some sort of singularity occurring at $\mu\mu_0 = m^2$, that is an artifact of the particular presentation of the expressions. In reality, all of these counter-terms are continuous functions of $m$, $\mu$, and $\mu_0$ for all positive real values of the parameters. A plot of these counter-terms is presented in figure \ref{fig:3}, where it is evident that they are continuous around $\mu\mu_0 = m^2$.

\section{Beta functions} \label{sec:5}
Using the counter-terms calculated in the previous section, one can find the associated beta functions for the dimensionless parameters $m$, $\mu$, and $\lambda$. These beta functions can be computed from the Callan-Symanzik equation. For a dimensionless coupling constant $g$ associated to a vertex with $m$ scalars and $n$ fermions the beta function takes the general form \cite{Peskin}
\begin{equation}
    \beta(g) = \fr{\pa g}{\pa \log(\Lambda)} = 
    \Lambda \fr{\pa}{\pa \Lambda} \left(-\delta_g + \frac{n g}{2}  \delta_{Z_\psi}\right).
\end{equation}
Note that in this equation we took the scalar renormalization to be zero because it vanishes to all orders in perturbation theory, as was shown in section \ref{sec:counterterms}.

The 1-loop beta functions for the dimensionless couplings are
\begin{\eq} \label{eq:5.2} 
\begin{split}
 \beta(\lambda)&= \Lambda \fr{\pa}{\pa \Lambda} \bigg ( - \delta_\lambda + \fr{1}{2}\lambda (2 \delta_{Z_\psi}) \bigg ) = 0, \\
 \beta(1/m) &= \Lambda \fr{\pa}{\pa \Lambda} \bigg ( - \delta_{1/m} + \fr{1}{2}\fr{1}{m} (2 \delta_{Z_\psi}) \bigg ) = \Xi_2, \\
 \beta(1/\mu) &= \Lambda \fr{\pa}{\pa \Lambda} \bigg ( - \delta_{1/\mu} \bigg ) =  0.
\end{split}
\end{\eq}

Surprisingly, the beta function for $\lambda$ vanishes at 1-loop even though both counter-terms $\delta_\lambda$ and $\delta_{Z_\psi}$ are independent nontrivial functions of the couplings. Indeed this vanishing beta function arises from a hidden symmetry of the Lagrangian \eqref{MinkLagrangian} given by the transformations
\begin{\eq}\label{exoticsym}
\psi \rightarrow e^{i\lambda \alpha t}\psi, \; \psi^\da \rightarrow e^{-i\lambda \alpha t} \psi^\da,\; \phi \rightarrow \phi + \alpha xy .
\end{\eq}
Under this symmetry the Lagrangian is invariant up to a total derivative term\footnote{In (Minkowski signature) momentum space, this transformation reads
\begin{equation*}
    \begin{split}
        \phi(\vec{k},\omega)&\to \phi(\vec{k},\omega) -\alpha (2\pi)^3 \delta(\omega)\frac{\partial^2}{\partial k_x\partial k_y}\delta^{(2)}(\vec{k})\\
        \psi(\vec{k},\omega)&\to \psi(\vec{k},\omega -\lambda\alpha)\\
        \psi^\dagger(\vec{k},\omega)&\to \psi(\vec{k},\omega +\lambda\alpha)
    \end{split}
\end{equation*}
which is compatible with our hard cutoff on the $\vec{k}$ integration.
}.
This symmetry is preserved by our regularization scheme, so it remains an invariance of the full renormalized Lagrangian \eqref{eq:full_lag}. This implies that the counter terms $\delta_{Z_\psi}$ and $\delta_\lambda$ are related by
\begin{\eq}
\lambda \delta_{Z_\psi} = \delta_\lambda
\end{\eq}
Hence, as the scalar does not acquire an anomalous dimension at any order in perturbation theory, we must have $\beta(\lambda)=0$ to all orders in perturbation theory. Therefore $\lambda$ is scale invariant within this theory. This symmetry is reminiscent of the emergent gauge symmetry observed in certain non-fermi liquids \cite{Metlitski2010,Mross:2010rd,Lee:2017njh}, though in our case the emergent symmetry is not anomalous.

\begin{figure}[h] 
\begin{center}
\includegraphics[width = 12 cm]{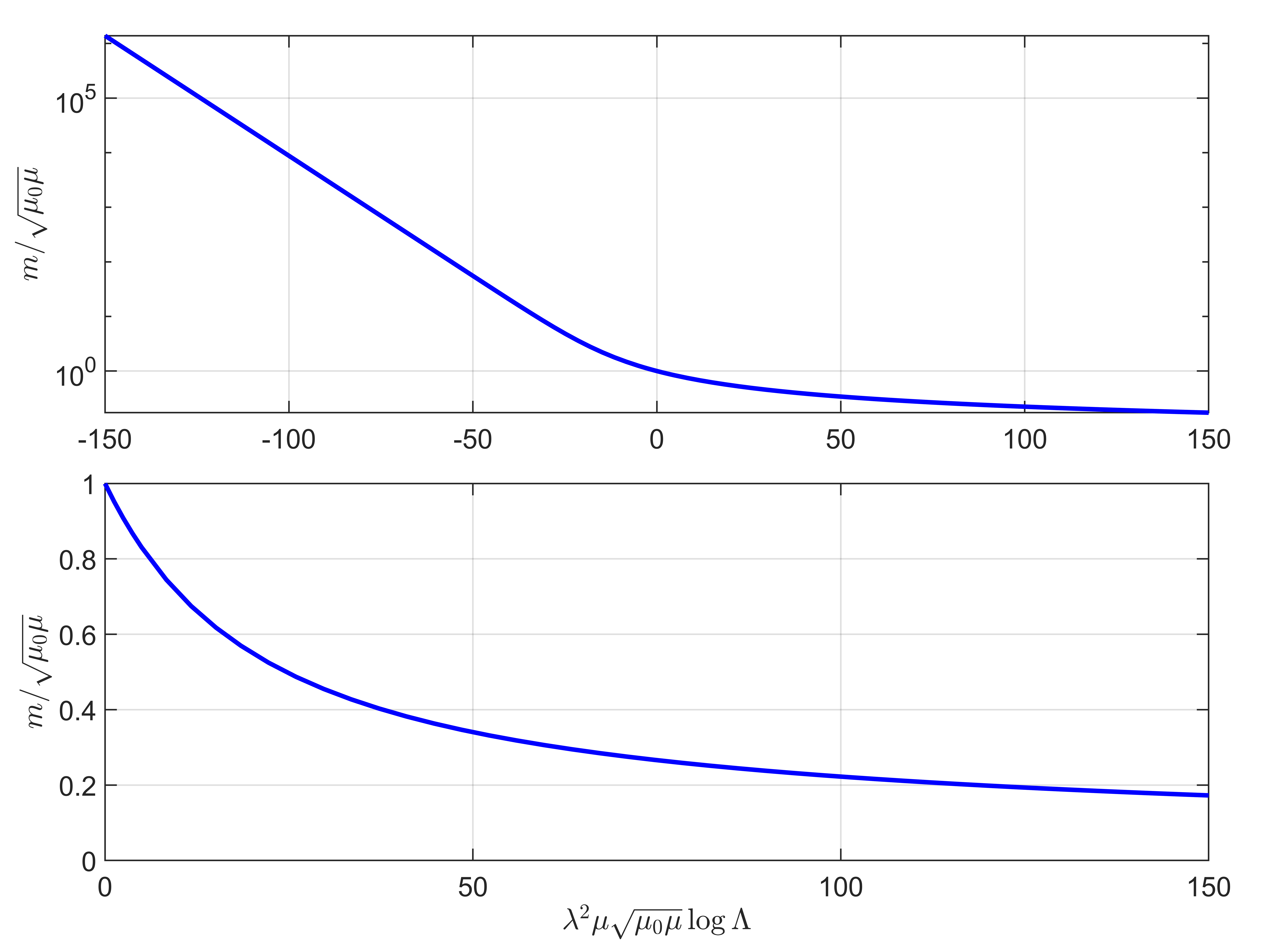}
\end{center}
\caption{ A plot of the dimensionless coupling $m/\sqrt{\mu_0\mu}$ as a function of the cutoff scale $\Lambda$, found by numerically solving the $\beta$ functions in equation \eqref{eq:5.2}. We present both a semi-log and a regular plot.}
\label{fig:5}
\end{figure}

As $\lambda$ does not run, we can take it to be uniformly small and analyze the RG flow of the remaining constant, $m$, just using its 1-loop $\beta$ function.  It is possible to consider a dimensionless version of this running by using the non-dimensional coupling $m/\sqrt{\mu_0 \mu}$, which completely removes any dependence on the specific constant values $\mu$ and $\mu_0$. The running of $m/\sqrt{\mu_0 \mu}$ to 1-loop order can be solved numerically, and a plot of this coupling constant as a function of the cutoff scale $\Lambda$ is presented in figure \ref{fig:5}. From the plot we see that in the deep IR, when $\Lambda \rightarrow \infty$, $m/\sqrt{\mu_0 \mu}$ very slowly approaches zero. On the other hand, $m/\sqrt{\mu_0 \mu}$ grows very fast in the UV, when $\Lambda \rightarrow -\infty$.


\section{Finite non-zero temperature}\label{sec:7}  
Another interesting dynamic of the theory occurs at non-zero temperature. In this situation, Euclidean time becomes periodic with periodicity $\beta$, where $\fr{1}{\beta}$ represents the temperature of the associated QFT. Due to the compactification the integral over $\omega$ will become a sum over Matsubara modes
\begin{\eq}
\fr{1}{2\pi} \int_{-\infty}^\infty d\omega \rightarrow \fr{1}{\beta} \sum_n
\end{\eq}
We can use this procedure to compute the perturbative 1-loop corrections also at finite temperatures, however non-perturbative effects may become important at  finite temperature. 

In particular our perturbative analysis was conducted at zero fermionic density, but this density is no longer zero at finite temperatures even for the non-interacting theory with $\lambda = 0$. We know that in 2-dimensions a non-interacting Fermi gas at a fixed chemical potential $\tilde{\mu}$,\footnote{Recall that the  chemical potential $\tilde{\mu} = -\gamma$ in our Lagrangian.} has a density $n$ given by
\begin{equation}
n =  \int_0^\infty \frac{dn}{d\epsilon}  \frac{1}{e^{\beta(\epsilon-\tilde{\mu})} + 1} d\epsilon
 = \frac{m}{2\pi} \frac{\log\left(1 + e^{\beta \tilde{\mu}} \right)}{\beta} .
\end{equation}
So at least when $\tilde{\mu} < 0$ and $\beta|\tilde{\mu}| \gg 1$ the density is exponentially small, and we can hope that the perturbative analysis is valid. We note that non-perturbative effects in the full interacting theory may still make this small density significant, but nevertheless we shall present the naive 1-loop perturbative calculation at finite temperature bellow.

\subsection{Scalar self-energy at finite temperature}

We shall start by computing the 1-loop scalar self energy at finite temperatures, which takes the form 
\begin{\eq} \label{eq:S1pI_finiteT}
\begin{split}
\Sigma_\phi  &=  \fr{\lambda^2k_x'^2k_y'^2}{(2\pi)^2} \fr{1}{\beta}\sum_{n = -\infty}^\infty \int d^2k \bigg ( \fr{1}{ +i(\fr{2\pi n}{\beta}+ \fr{\pi}{\beta}) + \fr{k^2}{2m}+\gamma} \bigg ) \bigg ( \fr{1}{-i(\omega'-(\fr{2\pi n}{\beta}+ \fr{\pi}{\beta})) + \fr{(k'-k)^2}{2m}+\gamma}  \bigg ) .
\end{split}
\end{\eq}
We can compute the sum over Matsubara modes by using the property that \footnote{ This is a consequence of using the residue theorem to evaluate the integral of $\cot(\pi z) f(z)$ on the contour $|z| = R \rightarrow \infty$, assuming that $|f(z)|$ decays sufficiently fast as $|z| \rightarrow \infty$ and that $f(z)$ has no poles on the integers.}
\begin{\eq}\label{eq:sum}
\sum_{n=-\infty}^{\infty} f(n) = - \sum_i \text{Res} \bigg ( \pi \cot(\pi z) f(z) , z_i \bigg ) , 
\end{\eq}
where the sum on the RHS runs only over the poles of $f(z)$. The summed expression in the scalar self-energy, \eqref{eq:S1pI_finiteT}, has two poles (as a function of $n$) located at
\begin{\eq}
z_1 = + i\fr{\beta}{2\pi} \bigg ( \fr{k^2}{2m} + \gamma \bigg ) - \fr{1}{2}, \qquad \qquad z_2 = i\fr{\beta}{2\pi} \bigg ( \fr{(k'-k)^2}{2m}+\gamma -i\omega'\bigg ) - \fr{1}{2}
\end{\eq}
This allows us to evaluate the sum using the residue theorem as 
\begin{\eq}
\begin{split}
\sum_{n=-\infty}^{\infty} f(n) = 
\frac{\beta}{2}\left[\tanh \bigg ( \fr{\beta}{2} \bigg ( \fr{k^2}{2m} + \gamma \bigg ) \bigg )
- \tanh \bigg ( \fr{\beta}{2} \bigg ( \fr{(k-k')^2}{2m} + \gamma - i \omega' \bigg ) \bigg )\right] \frac{1}{\fr{(k-k')^2 - k^2}{2m}  -i\omega'}.
\end{split} 
\end{\eq} 
Then the 1-loop scalar self energy at finite temperatures is
\begin{\eq} \label{eq:s1PI_finite_T_2}
\begin{split}
\Sigma_\phi = \fr{\lambda^2k_x'^2k_y'^2}{2(2\pi)^2} \int d^2k &\left[\tanh \bigg ( \fr{\beta}{2} \bigg ( \fr{k^2}{2m} + \gamma \bigg ) \bigg )
- \tanh \bigg ( \fr{\beta}{2} \bigg ( \fr{(k-k')^2}{2m} + \gamma - i \omega' \bigg ) \bigg )\right]\\
& \times \frac{1}{-i\omega' + \fr{ (k-k')^2 - k^2}{2m}  }.
\end{split} 
\end{\eq} 
Here $\omega'$ and $k'$ are the external Euclidean frequency and momentum. Note that when the external frequency $\omega'$ is evaluated on the Lorentzian energy, that is take $\omega' = -i\Omega$ where $\Omega$ is the Lorentzian energy, then these expressions become manifestly real.

This expression will reproduce the zero-temperature self-energy found in the previous section when we take the limit $\beta \rightarrow \infty$. This is due to the fact that the entire $\beta$ dependence of the self-energy arises from hyperbolic cotangent functions, which have the limiting behavior $\tanh(x) \approx 1 + O(e^{-2x})$ at large $x$. Additionally this integral is UV finite because of the same limiting behavior, only now we take $k^2 \rightarrow \infty$.

More generally in the perturbative picture the counter-terms for the finite QFT are consistent with the zero temperature QFT. Due to this, the $\beta$ function for $\lambda$ is also zero at finite temperatures. This is interesting as the symmetry that protects $\lambda$ from running is explicitly broken by the periodicity of Euclidean time, yet $\lambda$ does not run. This is because the symmetry is only softly broken as the UV physics is unaffected by the periodicity in Euclidean time.

\subsection{Fermion self-energy at finite temperature}

The fermion self-energy at finite temperature takes the form 
\begin{\eq} 
\Sigma_\psi =   \fr{\lambda^2}{\beta} \sum_{n = -\infty}^\infty\int \fr{d^2k}{(2\pi)^2} \fr{1}{-i(\omega_1 - (\fr{2\pi n}{\beta})) + \fr{(k_1 - k )^2}{2m} + \gamma}\fr{k_x^2k_y^2}{\mu_0 (\fr{2\pi n}{\beta})^2 + \fr{k_x^2k_y^2}{\mu}} .
\end{\eq}
One can again compute this sum using \eqref{eq:sum}, noting that there are three poles in the expression for the fermion self-energy: 
\begin{\eq}
z_{1,2} = \pm i\fr{\beta |k_xk_y|}{2\pi\sqrt{\mu\mu_0}}, \; \; z_3 = i\fr{\beta}{2\pi} \bigg [ -i \omega_1  + \fr{(k_1-k)^2}{2m}+\gamma  \bigg ] .
\end{\eq}
This leads to the sum of the form 
\begin{\eq}
\begin{split}
\sum_{n=-\infty}^{\infty} f(n) &= \frac{\beta  \sqrt{\fr{\mu}{\mu_0}} }{4} \coth \bigg (\fr{\beta |k_xk_y|}{2\sqrt{\mu\mu_0}} \bigg )  \fr{|k_xk_y|}{-i\omega_1 + \fr{|k_xk_y|}{\sqrt{\mu\mu_0}}  + \fr{(k_1 - k )^2}{2m} + \gamma}  \\ 
&+ \frac{\beta  \sqrt{\fr{\mu}{\mu_0}} }{4} \coth \bigg (\fr{\beta |k_xk_y|}{2\sqrt{\mu\mu_0}} \bigg )  \fr{|k_xk_y|}{-i\omega_1 - \fr{|k_xk_y|}{\sqrt{\mu\mu_0}}  + \fr{(k_1 - k )^2}{2m} + \gamma}  \\ 
&+ \fr{\beta}{2} \coth \bigg ( \fr{\beta}{2} \bigg [ \fr{(k_1-k)^2}{2m}+\gamma -i\omega_1 \bigg ] \bigg )  \fr{k_x^2k_y^2}{-\mu_0  \bigg ( \fr{(k_1-k)^2}{2m}+\gamma - i \omega_1\bigg )^2 + \fr{k_x^2k_y^2}{\mu}} .
\end{split}
\end{\eq}

We can combine these into the single compact expression
\begin{\eq}
\begin{split}
\sum_{n=-\infty}^{\infty} f(n) &= \frac{\beta  \mu }{2} \left[\coth \bigg (\fr{\beta \Upsilon}{2}\bigg )\Delta - \coth\bigg(\fr{\beta \Delta}{2}\bigg )\Upsilon  \right] \fr{\Upsilon}{\Delta^2 - \Upsilon^2} ,
\end{split}
\end{\eq}
where
\begin{equation}
    \Upsilon = \frac{|k_x k_y|}{\sqrt{\mu_0\mu}}, \qquad \qquad 
    \Delta = \fr{(k_1-k)^2}{2m}+\gamma -i\omega_1.
\end{equation}
As before, note that $\Delta$, and thus the whole expression, is real when considering an external Lorentzian energy $\omega_1 = -i \Omega_1$.

All together, leads to the fermion self-energy becoming
\begin{\eq}
\Sigma_\psi =  \frac{\lambda^2 \mu }{2} \int \fr{d^2k}{(2\pi)^2}  \left[\coth \bigg (\fr{\beta \Upsilon}{2}\bigg )\Delta - \coth\bigg(\fr{\beta \Delta}{2}\bigg )\Upsilon  \right] \fr{\Upsilon}{\Delta^2 - \Upsilon^2} .
\end{\eq}

Note that the integrand is finite when $\Delta = \Upsilon$, and reduces to the zero temperature expression \eqref{eq:27} in the limit $\beta \rightarrow \infty$.

\subsection{Vertex correction at finite temperature}

The vertex correction at finite temperatures take the form
\begin{\eq}
\Gamma = - \fr{\lambda^3}{\beta} \sum_{n = -\infty}^\infty \int \fr{d^2k}{(2\pi)^2} \fr{1}{-i\omega_1 +i\fr{2\pi n}{\beta} + \fr{(k_1 - k )^2}{2m} + \gamma} \fr{k_x^2k_y^2}{\mu_0 (\fr{2\pi n}{\beta})^2 + \fr{k_x^2k_y^2}{\mu}}\fr{(k_1+k_2)_x(k_1+k_2)_y}{i\omega_2 + i\fr{2\pi n}{\beta} + \fr{(k_2 + k )^2}{2m} + \gamma} .
\end{\eq}

For simplicity we define the quantities
\begin{equation}
    \Upsilon = \frac{|k_x k_y|}{\sqrt{\mu_0\mu}}, \qquad 
    \Delta_1 = \fr{(k_1-k)^2}{2m}+\gamma -i\omega_1, \qquad 
    \Delta_2 = \fr{(k_2+k)^2}{2m}+\gamma +i\omega_2,
\end{equation}
so that the he vertex correction becomes
\begin{\eq}
\Gamma = - \fr{\mu \lambda^3}{\beta} (k_1+k_2)_x(k_1+k_2)_y \sum_{n = -\infty}^\infty \int \fr{d^2k}{(2\pi)^2}  \fr{\Upsilon^2}{(\fr{2\pi n}{\beta})^2 + \Upsilon^2 } \fr{1}{\left(\Delta_1 +i\fr{2\pi n}{\beta}\right)\left(\Delta_2 +i\fr{2\pi n}{\beta}\right)} .
\end{\eq}
As before, note that $\Delta_{1,2}$ are real when considering external Lorentzian energies $\omega_{1,2} = -i\Omega_{1,2}$.

We can then evaluate the sum over Matsubara modes using \eqref{eq:sum}, noting that the summed function has poles at 
\begin{\eq}
z_{1,2} = i\fr{\beta}{2\pi}\Delta_{1,2}, \qquad \qquad
z_{3,4} = \pm i\fr{\beta}{2\pi}\Upsilon.
\end{\eq}

This results in
\begin{equation}
    \begin{aligned}
        \sum_{n = -\infty}^\infty  \fr{1}{(\fr{2\pi n}{\beta})^2 + \Upsilon^2 } \fr{1}{\left(\Delta_1 +i\fr{2\pi n}{\beta}\right)\left(\Delta_2 +i\fr{2\pi n}{\beta}\right)} =&
        \frac{\beta}{2}\coth\left(\frac{\beta}{2}\Delta_1 \right) \frac{1}{\left(\Upsilon^2 - \Delta_1^2 \right)\left(\Delta_2 - \Delta_1\right)}\\
        &+ \frac{\beta}{2}\coth\left(\frac{\beta}{2}\Delta_2 \right) \frac{1}{\left(\Upsilon^2 - \Delta_2^2 \right)\left(\Delta_1 - \Delta_2\right)}\\
        &+ \frac{\beta}{4}\coth\left(\frac{\beta}{2}\Upsilon \right) \frac{1}{\Upsilon\left(\Upsilon - \Delta_1 \right)\left(\Upsilon - \Delta_2 \right)}\\
        &+ \frac{\beta}{4}\coth\left(\frac{\beta}{2}\Upsilon \right) \frac{1}{\Upsilon\left(\Upsilon + \Delta_1 \right)\left(\Upsilon + \Delta_2 \right)}.
    \end{aligned}
\end{equation}

Combining everything together, the 1-loop vertex correction at finite temperatures is
\begin{equation}
\begin{aligned}
    \Gamma = - \fr{\mu \lambda^3}{2}  \int \fr{d^2k}{(2\pi)^2} & \fr{(k_1+k_2)_x(k_1+k_2)_y~\Upsilon}{\left(\Upsilon^2 - \Delta_1^2 \right)\left(\Upsilon^2 - \Delta_2^2 \right)\left(\Delta_1 - \Delta_2\right)} \Bigg[ \coth\left(\frac{\beta\Upsilon}{2} \right)\left(\Delta_1\Delta_2+\Upsilon^2 \right) \left(\Delta_1 - \Delta_2 \right)\\
    &\qquad \qquad \qquad- \coth\left(\frac{\beta\Delta_1}{2} \right) \Upsilon\left(\Upsilon^2 -\Delta_2^2\right)
    +\coth\left(\frac{\beta\Delta_2}{2} \right) \Upsilon\left(\Upsilon^2 -\Delta_1^2\right)
    \Bigg].
\end{aligned}
\end{equation}
In this from it is clear that the resulting integral is well defined around $\Upsilon = \Delta_{1,2}$ and $\Delta_1 = \Delta_2$. 

As in the previous 1-loop calculations, $\Gamma$ reduces to the zero temperature integral \eqref{eq:white_rice} in the $\beta \rightarrow\infty$ limit, and the UV divergence is unaffected by the finite temperature. To better understand the impact of low-temperature physics on this theory, a numerical evaluation of the respective integral while subtracting off the associated counter-term \eqref{eq:4.10} would provide novel insight into the dynamics of the theory at finite temperature.

\section{The theory at finite density} \label{sec:8}

Up to now we have analyzed the theory when the average fermion density is zero. However many interesting phenomena occur at finite fermion densities. To induce a fermionic density we can  take the chemical potential to be positive. When this happens much our of our perturbative analysis breaks down as we are expanding around the wrong vacuum. 

The finite density of fermions leads to two interesting phenomena. The first is that the fermionic density induces a background configuration for the scalar field. The second is that at finite densities we must expand the low energy action around the Fermi surface to get an effective theory. In this effective description the scalar's unique dispersion relation results in unique dynamics and RG flow.

\subsection{Semi-classical approximation of the ground state configuration}

Considering the model with a finite positive chemical potential ($\gamma < 0$), we would like to understand how the finite fermion density effects the optimal scalar field configuration, at least at a semi-classical level. To find the optimal scalar configuration we will minimize the energy density of the ground state of the fermions subject to a fixed scalar field configuration. Thus we will treat the fermions as an ideal Fermi gas, while the scalars will be treated classically.

It is clear that time varying configurations of the scalar field are always suppressed, however spatially varying configurations may not be suppressed, as they can alter the effective chemical potential of the fermions. In particular we can restrict ourselves to configuration where $\partial_x \partial_y \phi$ is a constant, as such configuration cause a global shift to the chemical potential of the fermions, rather than spatially varying shifts which will be additionally suppressed by the fermionic kinetic term. Such configurations take the form $\phi = \alpha x y$ for some fixed value $\alpha$, up to a subsystem symmetry transformation.

This scalar configuration modifies the chemical potential for the fermions to 
\begin{\eq}
\tilde{\mu} = -\gamma - \alpha \lambda,
\end{\eq}
resulting in a shift to the fermionic ground state energy density. However such a configuration also adds an energy density of $\alpha^2/(2\mu)$ due to the scalar kinetic term. 

A free non-relativistic fermionic field with mass $m$, at zero temperature and subject to a chemical potential $\tilde{\mu}$ (which is also the Fermi energy) in 2 spatial dimensions has a ground energy density of
\begin{equation}
    \mathcal{E}_\text{fermions} = - \fr{m}{2\pi} \tilde{\mu}^2 \Theta(\tilde{\mu}),
\end{equation}
where $\Theta(x)$ is the Heaviside step function. Note that this energy density is negative as states bellow the Fermi surface have an energy less than the Fermi energy (which is the chemical potential), and so gives an overall negative contribution to the energy density. \footnote{ In general dimensions the density of states of a free ideal (spin-less) Fermi gas is \cite{jab}
\begin{\eq}
\fr{dn}{d\ep} = \fr{m^{d/2}d}{2^{d/2}\pi^{d/2}\Gamma(1+\fr{d}{2})}\ep^{\fr{d -2}{2} } .
\end{\eq}
Assuming the fermions occupy all states up the Fermi energy, the contribution to the energy density from these fermions is $\int_0^{\tilde{\mu}} \ep \fr{dn}{d\ep} d\ep$. However, each fermion also contributes a negative energy of $-\tilde{\mu}$ due to the chemical potential, so we must add $-n \tilde{\mu}$ to the fermionic energy density contribution, resulting in
\begin{equation}
    \mathcal{E}_{\text{fermions}} = \int_0^{\tilde{\mu}} \left(\ep - \tilde{\mu}\right)\fr{dn}{d\ep} d\ep  
    = -\fr{m^{d/2}}{2^{d/2-2}(d+2)\pi^{d/2}\Gamma(1+\fr{d}{2})}\tilde{\mu}^{\fr{d + 2}{2}} .
\end{equation}
}

The total energy density of this ground state configuration is 
\begin{equation}
    \mathcal{E} = \frac{\alpha^2}{2\mu} - \fr{m}{2\pi} \left( \gamma + \alpha \lambda\right)^2 \Theta(-\gamma - \alpha \lambda) .
\end{equation}

For $\gamma \geq 0$ we see that the configuration with $\alpha = 0$ is stable (assuming $\lambda$ is small,) indicating that our previous perturbative analysis at zero density and around $\phi = 0$ is valid. 

When $\gamma< 0$ the minimal energy configuration of the system is no longer at $\alpha = 0$, but rather at
\begin{equation}
    \alpha_{\min} =\frac{ \lambda \gamma m \mu}{\pi - m \mu\lambda^2 } .
\end{equation}

It is interesting to note that there is an instability that happens when $ \lambda^2 > \pi/(m \mu) $, irregardless of the sign of the chemical potential. It is also not clear to us what the correct effective description of the system is in this case, or if one even exists, though this does occur at strong coupling when our perturbative understanding breaks down. 

Finally, we note that even though this scalar configuration seems to spontaneously break translational symmetry, there is a combination of translations and subsystem symmetry that acts trivially on these configurations.

\subsection{The effective description at finite densities}

At finite densities the low energy fermionic excitation's lie near the Fermi surface. We would like to understand the effective description for these low-lying modes in the presence of the non-standard scalar field. In the typical picture of an interacting Fermi liquid, if we impose an effective energy cutoff $\Lambda \ll E_f$, then fermions that reside in far away patches near the Fermi surface cannot interact via a scalar field unless they are antipodal. This is because the intermediate scalar field would have energy of order $E_f \gg \Lambda$, and so would be integrated out in the effective description \cite{Shankar1994,polchinski1992}. However, in our system this is no longer the case due to the unique dispersion relation of the mediating scalar field. 

\begin{figure}[h] 
\begin{center}
\includegraphics[width = 6 cm]{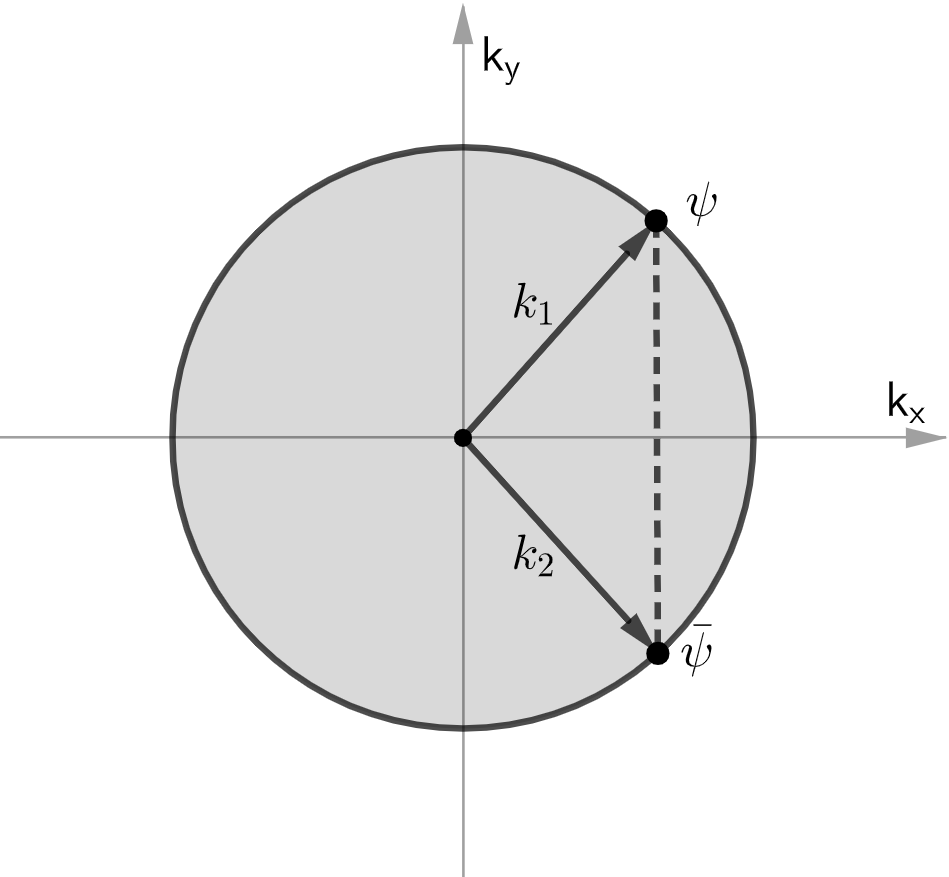}
\end{center}
\caption{A schematic sketch of the Fermi surface. All the states in the shaded region are occupied, while the effective description focuses on the fermions near the surface. The two fermions near the Fermi surface with momenta $k_1$ and $k_2$ can still interact even though $k_1 + k_2$ is large because $k_1 + k_2$ lies near the $k_x$ axis.}
\label{fig:fermi_surf}
\end{figure}

As the energy of the scalar field goes like $E \sim k_x^2 k_y^2$, the mediating scalar can have a small energy even with a large momenta $k_x$ in the $x$ direction, so long as $k_y$ is sufficiently small (or vice versa). Thus non antipodal fermions with momenta near the Fermi surface can still interact via the scalar field so long as they share the same momenta in $x$ or $y$ direction. An example of such fermions is presented in figure \ref{fig:fermi_surf}.

This unique interaction necessitates a four patch effective description for the fermions, rather than the more standard two patch description. Additionally, the strength of the coupling can depend on which specific four patches we focus on. Even in a single four patch description the strength of the coupling between the different patches differs. All these unusual features may give rise to unique scaling behaviors, and novel physics, though thoroughly analyzing this effective description is beyond the scope of this paper.


\section{Conclusions} \label{sec:9} 
In this paper, we studied the properties of a continuum field theory of a  fracton scalar interacting with a non-relativistic fermion. We properly renormalized the theory using a conventional ultraviolet cutoff scheme on momentum. From this, the associated 1-loop beta functions for the coupling constants were computed. We also explored some aspects of the theory at finite temperature and finite density.

The theory we studied may well be the simplest interacting theory with subsystem symmetries, in that it only allows a single marginal and no relevant interactions. Subsystem symmetries strongly restrict the allowed terms one can write down in the Lagrangian. For a single real scalar, no relevant or marginal terms exist that aren't quadratic in the fields. So our fracton plus fermion theory is a very natural first step in exploring interacting theories with subsystem symmetry.

There are several lessons one can draw from this exercise. Maybe first of all, it is reassuring to see that in this system, despite all of its peculiarities, including discontinuous field configurations and strong UV/IR mixing, the standard tools of quantum field theory still apply. As we discussed, one could wonder whether different RG schemes could be tried to extract a different scaling regime, but the standard momentum cutoff we employed does give meaningful results and, as we mentioned in the text, the fact that we couple the scalar with its unusual dispersion relations to a conventional fermion also seems to indicate that this is the correct scheme to use in this context.

Maybe the biggest surprise we encountered was the vanishing of the 1-loop beta function for the coupling constant. We have identified a novel symmetry \eqref{exoticsym} in the system that seem to guarantee this vanishing to all loops. Nevertheless, given the novelty of the subject it may be re-assuring to explicitly check in the future that the vanishing does indeed persist at 2-loop order. Beyond that, the kind of symmetry we observed, combining a time dependent phase rotation of a fermion with a shift in a scalar field, isn't unique to our system and, in fact, doesn't even rely on the presence of subsystem symmetries. A version of this symmetry can easily be constructed in more conventional systems of non-relativistic\footnote{In relativistic theories the fact that the kinetic terms involve a fermion bi-linear in the vector representation of the Lorentz group, while the Yukawa couplings involve a scalar bi-linear make it more challenging to find such symmetries in that context} fermions coupled to scalars. In particular, one can easily see that such a symmetry is in fact present in the theory of a critical metal, a Fermi surface coupled to an order parameter. This theory has been intensely studied over the years as for example reviewed in \cite{Lee:2017njh}. Some aspects of this symmetry have been discussed in \cite{Metlitski2010} and, in more detail, in appendix A of \cite{Mross:2010rd}. It would be very interesting to see if one could use this symmetry to constrain this system further, maybe along the lines we explored in here.

Our model has a global $U(1)$ symmetry, under which $\psi$ and $\psi^\dagger$ rotate with opposite phases (and $\phi$ is invariant). One obvious generalization would be to gauge this symmetry by introducing a dynamical gauge field $A_\mu(\vec{x},t)$, such that the gauge symmetry is
\begin{equation*}
\psi\to e^{i\theta(\vec{x},t)}\psi,\qquad\psi^\dagger\to e^{-i\theta(\vec{x},t)}\psi^\dagger,\qquad A_\mu\to A_\mu +\partial_\mu\theta(\vec{x},t),
\end{equation*}
and promoting derivatives to covariant derivatives. The Lagrangian \eqref{MinkLagrangian} becomes
\begin{equation}\label{MinkLagrangianGauged}
\begin{split}
\mathcal{L} &= \frac{\mu_0}{2}(\partial_0\phi)^2 - \frac{1}{2\mu}(\partial_x\partial_y \phi)^2  + \psi^\dagger \Bigl( i(\partial_0-iA_0) +  \frac{(\vec{\nabla}-i\vec{A})\cdot(\vec{\nabla}-i\vec{A})}{2m} \Bigr) \psi \\
&-\frac{1}{4e^2}F_{\mu\nu}F^{\mu\nu}+ \lambda \psi^\dagger \psi \partial_x\partial_y \phi - \gamma \psi^\dagger \psi .
\end{split}
\end{equation}
One might worry that, in the presence of a nontrivial gauge field background,  our symmetry \eqref{exoticsym} might suffer from an anomaly.  Fortunately the anomaly coefficient, which is proportional to the scalar self energy, vanishes, and so \eqref{exoticsym} remains a symmetry of the quantum theory of \eqref{MinkLagrangianGauged}. 


A very interesting future study in our system would be to analyse loops in the theory at finite density. In this case we would be describing fermions near the Fermi surface interacting with the fracton scalar. This will give an interesting structure to the dispersion relations of both of our matter fields, allowing for the possibility of interesting new scaling behaviors and potentially new physics.

\acknowledgments

We would like to thank L. Iliesiu and K. Jensen for useful discussions. We would also like to thank D. B. Kaplan for collaboration during early stages of this work.  We would also like thank D. T. Son for pointing out an error in the previous version.  The work of AK and AR was supported, in part, by the U.S. Department of Energy under Grant No. DE-SC0022021 and a grant from the Simons Foundation (Grant 651440, AK). The work of JD was supported in part by the National Science Foundation under Grant No.~PHY--1914679. The work of MJ was supported in part by the Mary Gates Research Scholarship, the Washington Research Fellowhip, and the Thomas Sedlock Icon Scholarship.

\appendix

\section{Computation of the fermion self-energy} \label{A2}
In this appendix, we go through the explicit calculation of the Fermion self-energy counter-terms. 

This will consist of computing the integral from  
\eqref{eq:fermion_1PI}. The associated integral is
\begin{\eq} \label{eq:23}
\Sigma_\psi = (-\lambda)^2  \int \fr{d\omega d^2k}{(2\pi)^3} \fr{1}{-i(\omega_1 - \omega) + \fr{(k_1 - k )^2}{2m} + \gamma}\fr{k_x^2k_y^2}{\mu_0 \omega^2 + \fr{k_x^2k_y^2}{\mu}} .
\end{\eq}
This $\omega$ integral is computed by  closing the contour in the lower half plane, then using the residue theorem. This contour only encloses a single pole arising from the scalar propagator. This component is computed by solving the residue of the form 
\begin{\eq}\label{eq:24}
\begin{split}
\int \fr{d\omega}{2\pi} &\fr{1}{-i(\omega_1 - \omega) +  \fr{(k_1 - k )^2}{2m} + \gamma }\fr{k_x^2k_y^2}{\mu_0 \omega^2 + \fr{k_x^2k_y^2}{\mu}}  \\
&\qquad \qquad = - i\text{Res} \bigg (\fr{1}{-i(\omega_1 - \omega) +  \fr{(k_1 - k )^2}{2m} + \gamma}
\fr{k_x^2k_y^2}{\mu_0 \omega^2 + \fr{k_x^2k_y^2}{\mu}}, \omega = -\fr{i|k_xk_y|}{\sqrt{\mu\mu_0} }\bigg )\\
&\qquad \qquad = \fr{\sqrt{\fr{\mu}{\mu_0}}}{2} \fr{|k_xk_y|}{-i\omega_1 + \fr{|k_xk_y|}{\sqrt{\mu_0\mu}} + \fr{(k_1-k)^2}{2m} + \gamma}   .
\end{split}
\end{\eq}
This therefore means that \eqref{eq:23} reduces down to 
\begin{\eq}\label{eq:27}
\begin{split}
\Sigma_\psi  &=  (-\lambda)^2 \fr{\sqrt{\fr{\mu}{\mu_0}}}{2}  \int \fr{dk_xdk_y}{(2\pi)^2}  \fr{|k_xk_y|}{-i\omega_1 + \fr{|k_xk_y|}{\sqrt{\mu_0\mu}} + \fr{(k_{1,x}-k_x)^2+(k_{1,y} - k_y)^2}{2m} + \gamma} .
\end{split}
\end{\eq}
To calculate this integral, we can transition into polar coordinates: $r^2 \equiv k_x^2 + k_y^2$, $k_x \equiv r\cos(\theta)$, $k_y \equiv r\sin(\theta)$. Due to the fact that we are only concerned with finding the appropriate counter-terms for this diagram, we seek to isolate the UV divergent piece. In this case, we will find that this integral becomes 
\begin{\eq}
\Sigma_\psi =  (-\lambda)^2 \fr{\sqrt{\fr{\mu}{\mu_0}}}{2(2\pi)^2}  \int_0^{2\pi} d\theta \int_0^\Lambda dr \fr{r^3\lvert\cos(\theta)\sin(\theta)|}{-i\omega_1 +  \fr{r^2\lvert\cos(\theta)\sin(\theta)|}{\sqrt{\mu_0\mu}} + \fr{(k_{1,x} - r\cos(\theta))^2 +(k_{1,y} - r\sin(\theta))^2}{2m} + \gamma }  .
\end{\eq}
If we now choose to Laurent expand the integrand above in terms of $r$ (or $\Lambda$), then we find that the self-energy splits into two UV divergent integrals. The two integrals will correspond to counter-terms that are defined in \eqref{eq:4.4}. These two integrals are defined as the quardratically divergent piece and the lograithmically divergent piece,
\begin{\eq}
\Sigma_\psi = \Sigma_\psi^1  + \Sigma_\psi^2 + O(\Lambda^{0}) .
\end{\eq}
Expanding the integral out, we find that $\Sigma_\psi^1$ and $\Sigma_\psi^2$ are
\begin{\eq}
\begin{split}
\Sigma_\psi^1 &= \lambda^2 \fr{m\mu}{2(2\pi)^2} \int_0^{2\pi} d\theta \fr{  |\sin(2\theta)|}{\sqrt{\mu\mu_0}+m|\sin(2\theta)|}  \int_0^\Lambda dr \; r  ,
\end{split}
\end{\eq}
and 
\begin{\eq} \label{eq:b7} 
\begin{split}
\Sigma_\psi^2 = -\lambda^2 \fr{\sqrt{\fr{\mu}{\mu_0}}}{2(2\pi)^2}  \int_0^{2\pi} d\theta  \bigg [ \fr{m |\sin(2\theta)| [-k_1^2 + 2m\gamma - 2 im \omega_1 + m (k_1^2 + 2m\gamma - 2im\omega_1)|\fr{\sin(2\theta)}{\sqrt{\mu\mu_0}}|]}{(1 + m |\fr{\sin(2\theta)}{\sqrt{\mu\mu_0}}|)^3} \bigg ] \int^\Lambda dr \fr{1}{r} .
\end{split} 
\end{\eq}

Starting with $\Sigma_\psi^1$, computing the $\theta$ integral leads to
\begin{\eq}
\begin{split}
\Sigma_\psi^1 &= \lambda^2 \fr{\mu}{2\pi^2} \bigg [  \bigg ( \theta - \fr{2\sqrt{\mu\mu_0}\arctan(\fr{m+2\sqrt{\mu\mu_0}\tan(\theta)}{\sqrt{4\mu\mu_0-m^2}})}{\sqrt{4\mu\mu_0-m^2}} \bigg ) \bigg ]_0^{\pi/2} \int_0^\Lambda dr \; r  \\
&= \lambda^2 \fr{\mu }{2\pi^2} \left\{  \fr{\pi}{2} +  \sqrt{\fr{\mu\mu_0}{\mu\mu_0-m^2}}\left( \arctan \bigg ( \fr{m}{\sqrt{\mu\mu_0 -m^2}} \bigg ) - \fr{\pi}{2} \right)  \right\} \int_0^\Lambda dr \; r .
\end{split}
\end{\eq} 
From this we see that $\Xi_3$ in \eqref{eq:4.4}, which is the coefficient term proportional to $\fr{\Lambda^2}{2}$ in the self energy, is
\begin{\eq}
\begin{split}
\Xi_3 = \lambda^2 \fr{\mu }{2\pi^2} \left\{  \fr{\pi}{2} +  \sqrt{\fr{\mu\mu_0}{\mu\mu_0-m^2}}\left( \arctan \bigg ( \fr{m}{\sqrt{\mu\mu_0 -m^2}} \bigg ) - \fr{\pi}{2} \right)  \right\}
\end{split}.
\end{\eq}

We can now go on to compute $\Sigma_\psi^2$. In this case, the UV divergent integral possess two terms, one proportional to $(\fr{k_1^2}{2m} + \gamma - i\omega)$, and another term proportional to $k_1^2/2$, which in \eqref{eq:4.4} are defined as $\Xi_1$ and $\Xi_2$. These correspond to the UV divergent pieces contributing to the $\delta_{Z_\psi}$ and the $\delta_{1/m}$ counter-terms through \eqref{eq:4.5}. More specifically, we can say that 
\begin{\eq}
\Sigma_\psi^2 = \bigg (\fr{k_1^2}{2m} + \gamma - i\omega_1 \bigg ) \Xi_1 \log(\Lambda) + \frac{k_1^2}{2} \Xi_2 \log(\Lambda).
\end{\eq}
$\Xi_1$ and $\Xi_2$ are computed from \eqref{eq:b7} as
\begin{\eq}
\begin{split}
\Xi_1 &= -\lambda^2 \fr{m^2\sqrt{\fr{\mu}{\mu_0}}}{(2\pi)^2}  \int_0^{2\pi} d\theta  \bigg [ \fr{|\sin(2\theta)|}{(1 + m |\fr{\sin(2\theta)}{\sqrt{\mu\mu_0}}|)^2} \bigg ] ,\\ 
\Xi_2 &= \lambda^2 \fr{m\sqrt{\fr{\mu}{\mu_0}}}{2\pi^2}  \int_0^{2\pi} d\theta \bigg [ \fr{ |\sin(2\theta)| }{(1 + m |\fr{\sin(2\theta)}{\sqrt{\mu\mu_0}}|)^3} \bigg ] .
\end{split}
\end{\eq}
Evaluating theses two integrals directly, we get
\begin{\eq}
\begin{split}
\Xi_1 &=  -\lambda^2 \fr{m^2 \mu}{\pi^2} \frac{ \sqrt{\mu \mu_0}}{\mu\mu_0 - m^2}\left\{ 1 + \fr{m}{(\mu\mu_0-m^2)^{1/2}} \left( \arctan\left(\fr{m}{\sqrt{\mu \mu_0-m^2}}\right) - \fr{\pi}{2} \right)  \right\},\\
\Xi_2 &= \lambda^2 \fr{m \mu}{\pi^2}  \fr{(\mu \mu_0)^{3/2}}{(\mu\mu_0-m^2)^2} \left\{ 2 + \fr{m^2}{\mu\mu_0} + \fr{3m}{\sqrt{\mu\mu_0-m^2}} \bigg (\arctan \bigg (\fr{m}{\sqrt{\mu\mu_0-m^2}} \bigg )  - \fr{\pi}{2} \bigg ) \right\} .
\end{split}
\end{\eq}
This then become the desired coefficients for the counter-terms.

\section{Computation of vertex correction} \label{A3}

In this section, we compute the Feynman diagram for the vertex correction. The associated integral for this correction equals 
\begin{\eq} \label{eq:29}
\Gamma = (-\lambda)^3 \int \fr{d\omega d^2k}{(2\pi)^3} \fr{1}{-i(\omega_1 - \omega) + \fr{(k_1 - k )^2}{2m} + \gamma} \fr{k_x^2k_y^2}{\mu_0 \omega^2 + \fr{k_x^2k_y^2}{\mu}}\fr{(k_1+k_2)_x(k_1+k_2)_y}{i(\omega_2 + \omega) + \fr{(k_2 + k )^2}{2m} + \gamma} .
\end{\eq}
The $\omega$ integral can then be computed by the residue theorem. Specifically, we can close in the lower half plane, and compute one residue, as there only exits a single pole within that region. This gives
\begin{\eq}
\begin{split}
 \int &\fr{d\omega}{2\pi} \fr{1}{-i(\omega_1 - \omega) + \fr{(k_1 - k )^2}{2m} + \gamma} \fr{k_x^2k_y^2}{\mu_0 \omega^2 + \fr{k_x^2k_y^2}{\mu}}\fr{1}{i(\omega_2 + \omega) + \fr{(k_2 + k )^2}{2m} + \gamma}  
\\ &=- i\text{Res} \bigg (\fr{1}{-i(\omega_1 - \omega) + \fr{(k_1 - k )^2}{2m} + \gamma} \fr{k_x^2k_y^2}{\mu_0 \omega^2 + \fr{k_x^2k_y^2}{\mu}}\fr{1}{i(\omega_2 + \omega) + \fr{(k_2 + k )^2}{2m} + \gamma} , \omega = -\fr{i|k_xk_y|}{\sqrt{\mu\mu_0} }\bigg )
\\ &=   \fr{\sqrt{\fr{\mu}{\mu_0}}}{2}\fr{1}{-i\omega_1 + \fr{|k_xk_y|}{\sqrt{\mu_0\mu}} + \fr{(k_1-k)^2}{2m} + \gamma} \fr{|k_xk_y|}{i\omega_2 + \fr{|k_xk_y|}{\sqrt{\mu_0\mu}} + \fr{(k_1-k)^2}{2m} +  \gamma} .
\end{split}
\end{\eq}
We can now collect all of these terms together and substitute back into  \eqref{eq:29} to calculate the rest of the diagram, which becomes 
\begin{\eq} \label{eq:white_rice}
\Gamma = (-\lambda)^3 (k_1+k_2)_x(k_1+k_2)_y  \int \fr{d^2k}{(2\pi)^2} \fr{\sqrt{\fr{\mu}{\mu_0}}}{2}\fr{1}{-i\omega_1 + \fr{|k_xk_y|}{\sqrt{\mu_0\mu}} + \fr{(k_1-k)^2}{2m} + \gamma} \fr{|k_xk_y|}{i\omega_2 + \fr{|k_xk_y|}{\sqrt{\mu_0\mu}} + \fr{(k_1-k)^2}{2m} +  \gamma} .
\end{\eq}
To calculate this integral, we can transition into polar coordinates: $r^2 \equiv k_x^2 + k_y^2$, $k_x \equiv r\cos(\theta)$, $k_y \equiv r\sin(\theta)$. This leads to 
\begin{\eq}
\begin{split}
\Gamma = (-\lambda)^3 (k_1+k_2)_x(k_1+k_2)_y \bigg [\int &\fr{r dr d\theta}{(2\pi)^2} \fr{\sqrt{\fr{\mu}{\mu_0}}}{2}\fr{1}{-i\omega_1 + \fr{r^2|\cos(\theta)\sin(\theta)|}{\sqrt{\mu_0\mu}} + \fr{(k_1-k)^2}{2m} + \gamma} \\ &\times \fr{r^2|\cos(\theta)\sin(\theta)|}{i\omega_2 + \fr{r^2|\cos(\theta)\sin(\theta)|}{\sqrt{\mu_0\mu}} + \fr{(k_1-k)^2}{2m} +  \gamma} \bigg ] .
\end{split}
\end{\eq}
 As we are only interested in the UV divergent piece, we can expand the integral as a perturbative series in terms of external momenta. Only the leading order term within this series will contain the UV divergence, and this leading term is found by taking $k_1^\mu = k_2^\mu = (0,\vec{0})$ inside the integral. This then simplifies the vertex to become 
\begin{\eq}
\begin{split}
\Gamma_{\text{UV}} = (-\lambda)^3 (k_1+k_2)_x(k_1+k_2)_y \bigg [ \int &\fr{r dr d\theta}{(2\pi)^2} \fr{\sqrt{\fr{\mu}{\mu_0}}}{2}\fr{r^2|\cos(\theta)\sin(\theta)|}{ \left (  \fr{r^2|\cos(\theta)\sin(\theta)|}{\sqrt{\mu_0\mu}} + \fr{r^2}{2m} +  \gamma \right )^2 } \bigg ] \\ =  (-\lambda)^3 (k_1+k_2)_x(k_1+k_2)_y \Xi .
\end{split}
\end{\eq}
We can now compute the $\theta$ integral:
\begin{equation}
    \begin{aligned}
        \Xi &= \frac{\mu\sqrt{\mu \mu_0}}{\pi^2} \int_0^\Lambda \fr{dr}{r} \int_0^{\pi/2} d\theta \fr{\sin(2\theta)}{ \left (  \sin(2\theta)  + \fr{\sqrt{\mu_0\mu}}{m} +  \frac{2\sqrt{\mu_0\mu}\gamma}{r^2} \right )^2 } \\
        &= \frac{\mu\sqrt{\mu \mu_0}}{\pi^2} \int_0^\Lambda \fr{dr}{r}~ \fr{1}{ \left(\mu \mu_0 \left(\fr{1}{m} + \fr{2\gamma}{r^2} \right)^2-1\right)} \times\\
        & \qquad \qquad \qquad \times\left[1 + \fr{1}{\sqrt{\mu \mu_0 \left(\fr{1}{m} + \fr{2\gamma}{r^2} \right)^2-1}}\left( \arctan\left(\frac{1}{\sqrt{\mu \mu_0 \left(\fr{1}{m} + \fr{2\gamma}{r^2} \right)^2-1}} \right)- \frac{\pi}{2}\right)\right] .
    \end{aligned}
\end{equation}

Then the divergent piece of the $r$ integral is found by expanding the above function around $r \rightarrow \infty$, giving us
\begin{equation}
\Xi = \frac{\mu\sqrt{\mu \mu_0}}{\pi^2} \fr{m^2}{\mu \mu_0-m^2}
\left[1 + \fr{m}{\sqrt{\mu \mu_0 - m^2}}\left( \arctan\left(\frac{m}{\sqrt{\mu \mu_0-m^2}} \right)- \frac{\pi}{2}\right)\right] \log(\Lambda) + O(\Lambda^0).
\end{equation}

From this, we have that the associated counter-term for $\lambda$ is 
\begin{\eq}
\delta_\lambda = \lambda^3 \fr{m^2 \mu}{\pi^2} \fr{\sqrt{\mu\mu_0}}{\mu\mu_0 - m^2}
\left\{ 1 +\fr{m}{(\mu\mu_0-m^2)^{1/2}} \left( \arctan \left( \fr{m}{\sqrt{\mu\mu_0-m^2}} \right) - \fr{\pi}{2} \right)  \right\} \log(\Lambda) .
\end{\eq}

\section{Correlation functions} \label{sec:6} 
For this section, we will calculate the correction to the two point function for the scalar. Specifically, we will compute the $O(\lambda^2)$ correction to the two-point function. This will amount to calculating the integral of the form
\begin{\eq}
\langle \pa_x \pa_y \phi(\tau,x,y) \pa_x\pa_y \phi(0,0,0) \rangle = \fr{1}{(2\pi)^3} \int d\omega dk_x dk_y e^{-ixk_x - iyk_y -i\tau \omega}\fr{k_x^2k_y^2}{\mu_0 \omega^2 + \fr{k_x^2k_y^2}{\mu} + \Sigma_\phi} 
\end{\eq}
Here $\Sigma_\phi$ was computed in  and is given by
\begin{\eq}
\Sigma_\phi = \fr{m\lambda^2k_x^2k_y^2}{4\pi\mu_0}\log \left (\fr{k_x^2+k_y^2}{4} - m i\omega + 2m\gamma \right)
\end{\eq}
One can first compute the $\omega$ integral using the residue theorem. This equals
\begin{\eq} \label{eq:64}
\begin{split}
\int_{-\infty}^\infty \; d\omega  \; \fr{e^{-i\tau \omega}}{\mu_0 \omega^2 + \fr{k_x^2k_y^2}{\mu} + \fr{m\lambda^2k_x^2k_y^2}{4\pi}\log \left (\fr{k_x^2+k_y^2}{4} - m i\omega + 2m\gamma \right) } \\ = \fr{1}{\mu_0} \int_{-\infty}^\infty \; d\omega  \; \fr{e^{-i\tau \omega}}{\omega^2 + \fr{k_x^2k_y^2}{\mu\mu_0} + \fr{m\lambda^2k_x^2k_y^2}{4\pi\mu_0}\log \left (\fr{k_x^2+k_y^2}{4} - m i\omega + 2m\gamma \right) } 
\end{split}
\end{\eq}
Here we can extend the $\omega$ in terms of $\lambda$ in order to find the $\lambda^2$ correction to the pole. This particular integral only possess two poles, one in the upper and lower half planes, and when $\lambda =0$ the location of this poles $\omega = \pm \fr{i|k_xk_y|}{\sqrt{\mu\mu_0}}$. At small $\lambda$ the location of the poles, and the residue associated with the pole change perturbativley with $\lambda$. Thus, we can compute this integral perturbativley by computing the order by order change in the residue of the pole. The modified location of the pole is at $\omega^* = \omega_1 + \lambda^2 \omega_2 + O(\lambda^4)$. Here $\omega_1 = \fr{i|k_xk_y|}{\sqrt{\mu\mu_0}}$, and one can find $\omega_2$ by finding the term that leads the denominator of the integrand to equal zero to $O(\lambda^2)$. This leads to 
\begin{\eq}
\begin{split}
\omega_2 = -\fr{1}{2\omega_1} \fr{mk_x^2k_y^2}{4\pi\mu_0} \log \bigg ( \fr{k_x^2+k_y^2}{4} - m i\omega_1 + 2m\gamma\bigg ) = -\fr{\sqrt{\mu\mu_0}}{2i}\fr{m|k_xk_y|}{4\pi\mu_0} \log \bigg (  \fr{k_x^2+k_y^2}{4} +m\fr{|k_xk_y|}{\sqrt{\mu\mu_0}} + 2m\gamma\bigg ) 
\end{split}
\end{\eq}
One can now Taylor expand the integrand for \eqref{eq:64} around $\omega^* = \omega_1 + \lambda^2 \omega_2$ and simply take the leading order term. This leading order term will correspond to the residue of the integral. This equals 
\begin{\eq}
\fr{1}{\mu_0} \int_{-\infty}^\infty \; d\omega  \; \fr{e^{-i\tau \omega}}{\omega^2 + \fr{k_x^2k_y^2}{\mu\mu_0} + \fr{m\lambda^2k_x^2k_y^2}{4\pi\mu_0}\log \left (\fr{k_x^2+k_y^2}{4} - m i\omega + 2m\gamma \right) }  = \fr{2\pi }{\mu_0} \fr{e^{i\omega^*|\tau|}}{ 2\sqrt{\fr{k_x^2k_y^2}{\mu\mu_0} + \fr{m\lambda^2k_x^2k_y^2}{4\pi\mu_0}\log \left (\fr{k_x^2+k_y^2}{4} - m i\omega^* + 2m\gamma \right) }} 
\end{\eq}
In this last line we used the fact that the integral reduces down to the expression. We must now only take into account for terms $O(\lambda^2)$ within the square root. Thus, in this analysis, we use the fact that $\log(\omega^*) \approx \log(\omega_1)$. This means that the full two point function becomes 
\begin{\eq} \label{eq:6.7} 
\langle \pa_x \pa_y \phi(\tau,x,y) \pa_x\pa_y \phi(0,0,0) \rangle =  \fr{1}{(2\pi)^3}  \fr{2\pi }{\mu_0}\int dk_xdk_y e^{-ixk_x - iyk_y }\fr{k_x^2k_y^2e^{i\omega^*|\tau|}}{2\sqrt{\fr{k_x^2k_y^2}{\mu\mu_0} + \fr{m\lambda^2k_x^2k_y^2}{4\pi\mu_0}\log \left (\fr{k_x^2+k_y^2}{4} - m i\omega_1 + 2m\gamma \right) }}
\end{\eq}
One can now further simplify the integral above by taking into account its symmetry about the origin. This leads to  
\begin{\eq} \label{eq:D7}
\langle \pa_x \pa_y \phi(\tau,x,y) \pa_x\pa_y \phi(0,0,0) \rangle =\fr{1}{(2\pi)^3}  \fr{2\pi}{\mu_0}\int_{-\infty}^\infty dk_xdk_y \fr{|k_xk_y|\cos(k_xx)\cos(k_yy)e^{i\omega^*|\tau|}}{2\sqrt{\fr{1}{\mu\mu_0} + \fr{m\lambda^2}{4\pi\mu_0}\log \left (\fr{k_x^2+k_y^2}{4} + m\fr{|k_xk_y|}{\sqrt{\mu\mu_0}} + 2m\gamma \right) }}
\end{\eq}
In the case above, when the value of $\tau \rightarrow 0$ the integral \eqref{eq:D7} is no longer well defined. This is clear as the theory possess an interesting UV/IR mixing property. Specifically, the integral at non-zero $\tau$ leads to \eqref{eq:6.7}, however in the limit where $\tau \rightarrow 0$, the integral begins to diverge. Thus, as at smaller values of $\tau$, the correlation function begins to diverge. This gives rise to this apparent UV/IR mixing property. 

\bibliographystyle{JHEP}
\bibliography{references}

\end{document}